# Encoded packet-Assisted Rescue Approach to Reliable Unicast in Wireless Networks

Zhiheng Zhou, Liang Zhou, Xing Wang, Yuanquan Tan
National Key Laboratory of Communication
University of Electronic Science and Tech of China
Chengdu, China
E-mail: {zhzhou, lzhou, wangxing, tanyuanquan}@uestc.edu.cn

***Abstract*—Recently, there has been a growing interest of using network coding to support reliable unicast over an error-prone channel. However, previous network coding schemes focused only on native packets and ignored the role of encoded packets. In this paper, we develop an efficient retransmission approach with network coding method, namely Encoded packet-Assisted Rescue (EAR), which is able to overcome this limitation. Using the proposed network coding approach, clients store the overheard encoded packets and report to the sender. And then, when the sender makes coding decisions, it not only considers native packets, but also takes account of encoded packets. In this way, more coding opportunities are emerged, i.e. more packets can be mixed together, resulting in improving the retransmission efficiency. Moreover, theoretical analysis and simulation results show that comparing with the existing schemes, our schemes can greatly reduce the total number of retransmissions.**

***Keywords*--Error-prone; Network Coding; Unicast; Rescue;**

## I. INTRODUCTION

It is well-known that wireless networks are error-prone because of fading and interference. Many researches have revealed that IEEE 802.11-based wireless mesh networks suffer from severe packet loss radio in wide conditions [1]. Hence, automatic repeat request (ARQ) [2] technique is used to make a wireless link reliable and further improve end-to-end throughput. However, in traditional ARQ protocols, whenever a packet gets erased, the sender simply retransmits it, which usually consumes a considerable channel capacity. Then, Hybrid ARQ (HARQ) [3] protocols that implement Forward Error Coding (FEC) are introduced and have been proven to be more efficient than original ARQ protocols or pure FEC schemes under wide conditions [4].

Recently network coding (NC) [5] has emerged as a promising technique to increase the network bandwidth efficiency and reliability, since it enables to mix multiple incoming packets in a single transmission rather than just relaying these incoming packets to output links one by one, which breaks with the conventional store-and-forward way. Particularly, authors of COPE [6] presented a practical XOR-based network coding with the concept of opportuneistic listening and opportunistic coding for unicast in wireless mesh networks. It employed cumulative ACK in its COPE layer to provide reliable transmission. However, its performance is only sub-optimal in lossy wireless networks. The reason is that COPE does not consider the issue of how to effectively retransmit lost packets, which could cause its performance degradation under the high loss ratio. As an elegant reliable retransmission paradigm in one-to-many single-hop wireless networks, Nguyen et al. [8] combined network coding technique with the traditional ARQ technique, NC-ARQ, which can significantly enhance the bandwidth efficiency. Rozner et al. [9] modified the retransmission mechanism in COPE and proposed an efficient retransmission approach (ER). With ER, packets that need to be retransmitted are encoded together, such that multiple packet-loss for different destinations can be recovered by one retransmission. Since then, a lot of network coding based retransmission protocols [10-14] have been presented to provide efficient and reliable communications in lossy wireless networks.

Nonetheless, all of these protocols only focal on native packets that need to be retransmitted, and none consider the role of encoded packets, which contained these loss native packets and have been overheard by clients who aren't their destinations. Using the traditional network coding schemes, when clients overheard encoded packets, they just discard them or simply store them but don't report these information to the sender. Therefore, when the sender makes coding decisions, it can't utilize encoded packets to exploit coding opportunities, which is inefficient and expensive.

Therefore, we study an efficient retransmission approach with network coding technique, named Encoded packet-Assisted Rescue (EAR) in this paper, which is able to exploit the coding opportunities of both native and encoded packets. Our contribution is summarized as follows:

- We introduce a system model of the wireless one-to-many single-hop unicast system and then study the theoretical analysis in terms of the number of retransmissions of the proposed scheme.
- We define the *weight of packet-loss pattern*, which helps us to evaluate the EAR's performance more precisely and design a optimum coding algorithm to maximize the benefit of coded packets, namely to minimize the total number of retransmissions

The remainder of this paper is organized as follows. In next Section, we describe a system model in the context of wireless unicast scenario. In Section III, we present some theoretical derivations on the total number of retransmissions with the proposed EAR schemes. In Section IV, theory and simulation results are illustrated. Finally, conclusion is given in Section V.

## II. System model and assumptions

If multiple flows traverse the same node, we called such node *central node* (unless otherwise stated we simply call it node *C*) and there is the opportunity to apply network coding techniques to improve the overall retransmission efficiency for each flow crossed it. Consequently, we only consider the retransmissions between node *C* and its neighbors even for the wheel scenario in this paper. And we also consider that the throughput rate of each flow over the networks has already stabilized. Furthermore, we assume node *C* employs a sufficient large retransmission buffer to avoid too early rescue process. In particular, each receiver would request a distinct set of loss packets, which from node *C*'s point of view, corresponds to supporting different unicast sessions. Similar with [10], we make the following assumptions about the wireless unicast retransmission models.

- There are $N$ $(N > 2)$ receivers $R_i$ $(1 \leqslant i \leqslant N)$, and the central node retransmits lost packets after a fixed time slot $\Delta T$.
- We suppose route of each flow would pass through node *C* to maximize coding opportunities [7].
- Node *C* can always know the current packet-loss states of both native and encoded packets at all receivers. This can be carried through by using positive and negative acknowledgements (ACK/NAKs).
- To simplify the analysis, we assume all the feedback are instantaneous and reliable.
- Packet lost rates between node *C* and each receiver are mutually independent and follow the Bernoulli distribution, where each packet is lost with a fixed probability $\omega_i$ ($i$ is the receiver's ID, $1 \leqslant i \leqslant N$) at each receiver.

Let us consider the unicast topology, which consists of *N* receivers. Thereby, if each client requests *K* distinct packets, then node *C* needs to successfully deliver a total of $K \times N$ packets to all of them. To plainly represent the packet-loss state, we define packet-loss pattern as follow:

**Definition 1.** *Packet-loss pattern $\rho_P$ is a row vector that represents the current packet-loss state of packet P at all the receivers, thus its dimension is equal to the number of the receivers. When packet P is successfully obtained by receiver $R_i$, the $i^{th}$ entry in $\rho_P$ will be marked 1, or else 0.*

In this paper, the packet-loss pattern relating to the native packets is simply called *native pattern*, and *coded pattern* denotes the one relating to the encoding packets. The following definitions play crucial roles in this paper.

**Definition 2.** *The weight of a packet-loss pattern $W(\rho)$ is the number of non-zero elements in $\rho$. In particular, $W(\rho)$ $< N$ for N-receiver scenarios.*

**Remark:** Note that, if all the intended receivers of packet $P$ correctly receive it, there is no longer a packet-loss pattern relevant to packet $P$, vice versa.

Furthermore, node $C$ would retransmit $P$ several times to deliver it successfully to its nexthop, for the reason that the wireless channel is error-prone. And the packet-error pattern $\rho_P$ would be altered after each retransmission of $P$. The weight of the newer loss pattern (if it exists), however, is always no lighter than the previous one, in respect that the receivers who have overheard $P$ would reserve it until the end of this retransmission process.

At last, there are no less than one lost packet relating to a loss pattern. Thereby, unless otherwise stated, we relate a loss pattern to a set of packets which have the same loss state. Then, we define $P_\rho$ as the set of lost packets which have the same loss pattern $\rho$ during the rescue process. Thus, in retransmission process, if a non-empty set $P_\rho$ has been already transferred to the empty set, i.e. no loss packets relevant to $\rho$, we state that the central node has *rescued* the loss pattern $\rho$ or the set $P_\rho$. As we mentioned above, a packet belonging to $P_\rho$ is eliminated from $P_\rho$ due to the following two reasons:

*1)* The intended receiver of this packet has received it.

*2)* The intended receiver is failure to receive it again, yet the corresponding loss pattern is changed.

To sum up, we present the following theorem.

**Theorem 1.** *The expect number of retransmissions $\Omega_\rho$ requested by the central node to rescue loss-pattern $\rho$, whose the $i_r^{th}$ $(i_r \in N, r = 1, \dots, N - W(\rho))$ entries are equal to $0$, for N-receiver scenario is*

$$\Omega_\rho = \frac{|Or|}{1 - \prod \omega_{i_r}} \tag{1}$$

*where $Or$ is the set consisting of the lost packets which have the same loss pattern $\rho$ after the original transmission. And $\Omega_{\rho \to \rho'}$ packets in $P_\rho$ would be transferred to $P_{\rho'}$ after the central node rescued $P_\rho$.*

$$\Omega_{\rho \to \rho'} = \Omega_\rho \times Pr\{\rho' | \rho\} \tag{2}$$

*where $Pr\{\rho'|\rho\}$ denotes the probability that a packet is transferred from $P_\rho$ to $P_{\rho'}$.*

**Proof:** To simplify the analysis, we suppose that node $C$ would retransmits $P_\rho$ *round by round*, which means node $C$ first transmits the entire packets belonging to $P_\rho$ one by one, and collects their receive-state to modify $P_\rho$, and then it repeats the these steps for the residual packets in $P_\rho$ again and again until these is no longer a packet relating to $\rho$. We define $P_\rho^{(k)}(k > 0)$ as the set relevant to $\rho$ after the $k^{th}$ round, and $P_\rho^{(0)} = Or$. Let random variable $Y_k$ $(k > 0)$ represent the cardinality of $P_\rho^{(k)}$, and we have $Y_0 = |Or|$. For the reason that the deliveries are i.i.d. and follow the Bernoulli distribution, the random variables $Y_k$ $(k = 0, 1, \dots)$ are i.i.d. and follow the binomial distribution. Further-more, a packet is held in $P_\rho^{(k)}$ after the $k^{th}$ round, if and only if the receivers $R_{i_r}$ who lost it before are still failure to receive it. Hence, we have

$$E[Y_k] = E[E[Y_k|Y_{k-1}]] \quad = Y_0 \times (\prod \omega_{i_r})^k \tag{3}$$

due to $\prod \omega_{i_k} \leqslant 1$, the series $E[Y_k]\,(k > 0)$ is convergent, the

expect number of retransmissions $\Omega_\rho$, that node needs $C$ to rescues $\rho$, is

$$\begin{aligned}\Omega_\rho &= Y_0 + \sum\nolimits_{k=1}^{\infty} E[Y_k] \\ &= Y_0 \cdot (Pr\{\rho\})^0 + \sum\nolimits_{k=1}^{\infty} Y_0 \cdot (\prod \omega_{i_r})^k \\ &= Y_0 \cdot \sum\nolimits_{k=0}^{\infty} (\prod \omega_{i_r})^k = \frac{|Or|}{1 - \prod \omega_{i_r}} \qquad (4)\end{aligned}$$

Now, let random variable $Y_k'$ $(k > 0)$ represent the number of packets that are transferred from $P_\rho$ to $P_{\rho'}$, $\rho \neq \rho'$, after the $k^{th}$ round, particularly, $Y_0' = 0$. Random variables $Y_k'$ $(k > 0)$ are i.i.d as well and follow the binomial distribution. Then, we have

$$\begin{aligned}E[Y_k'] &= E[E[Y_k'|Y_{k-1}]] \\ &= E[Y_{k-1}] \times Pr\{\rho'|\rho\} \\ &= Y_0 \times (\prod \omega_{i_r})^{k-1} \times Pr\{\rho'|\rho\} \qquad (5)\end{aligned}$$

due to $\prod \omega_{i_r} \leqslant 1$, the series $E[Y_k']$ $(k > 0)$ is convergent, the number of packets that are transferred from $P_\rho$ to $P_{\rho'}$ after node $C$ has rescued $\rho$ is

$$\begin{aligned}\Omega_{\rho \to \rho'} &= \sum\nolimits_{k=1}^{\infty} E[Y_k'] \\ &= \sum\nolimits_{k=1}^{\infty} Y_0 \cdot (\prod \omega_{i_r})^{(k-1)} \cdot Pr\{\rho'|\rho\} \qquad (6) \\ &= \Omega_\rho \times Pr\{\rho'|\rho\}\end{aligned}$$

that packets in $P_\rho$ can be relocated to $P_{\rho'}$. ∎

Obviously, this theorem is suitable for both native and coded patterns. Finally, we introduce the following concept.

**Definition 3.** *If n packets respectively belonging to the sets $P_{\rho_1}, P_{\rho_2}, \ldots, P_{\rho_n}$ $(2 \leqslant n \leqslant N)$ can be coded together, we state the sets $P_{\rho_1}, P_{\rho_2}, \ldots, P_{\rho_n}$ can be encoded together, and we also say that the corresponding loss patterns $\rho_1$,…, $\rho_n$ can be encoded together.*

## III. Performance Analysis

In this section, we first review the available work on applying network coding to reliable multi-unicast. Then, we study some theoretical analysis in terms of total number of retransmissions of the traditional and the proposed EAR techniques for both wheel and single-hop scenarios. Before delving into details, we refer the reader to the following symbols, which be used in the rest of the paper.

- $P^j$: The set of native packets node $R_j$ requests.
- $\rho^j$: The set of all possible loss patterns relating to the native packets node $R_j$ requests.
- $\rho^{j_1,\ldots,j_n}$: The set of all possible coded patterns which are in relation to coded lost packets, each of which contains $n$ packets nodes $R_{j_1}, \ldots, R_{j_n}$ requested separately, where $j_k \in N, 1 \leqslant k \leqslant n, 2 \leqslant n \leqslant N$.
- $P_{\rho_k}^j$: the set of native lost packets which are reque-sted by node $R_j$ and have the loss state $\rho_k = \rho_k^j$.
- $P_{\rho_k}^{j_1,\ldots,j_n}$: the set of native lost packets have the loss state $\rho_k = \rho_k^{j_1,\ldots,j_n}$.

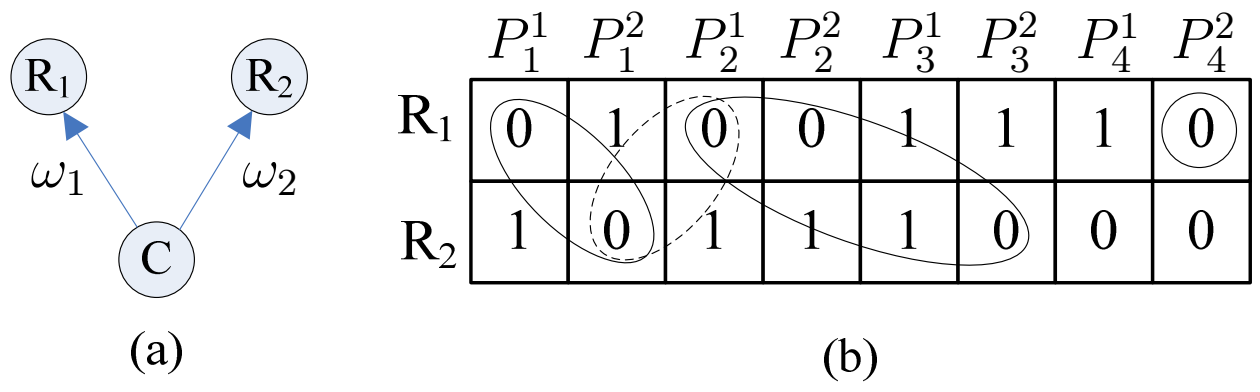

Figure 1. Example of NC-ARQ. (a) Two receivers uncast scenario. (b) Possible retransmission process: $P_1^1 \oplus P_1^2$, $P_2^1 \oplus P_3^2$, $P_4^2$ and $P_1^2 \oplus P_2^1$.

We also use notation $\rho_{Pj}$ to denote the set of loss patterns relating to packets in $P^j$. However, when $|P^j|$ is sufficiently large, we have $\rho_{Pj} = \rho^j$ because of the law of large numbers. Thus, unless otherwise stated, we also utilize $\rho^j$ to denote $\rho_{Pj}$. And it is the same to $\rho_{Pj_1,\ldots,j_n}$.

### A. Overview

The key idea of Nguyen et al. [8] and Rozner et al. [9] proposed network coding schemes is to first buffer the lost packets in the retransmission buffer during some time, then, instead of transmitting them one by one, the source combines the maximum number of lost packets with distinct intended receivers into one packet and delivers this coded packet in one retransmission.

To illustrate how this coding-base reliable retransmission scheme works, we take an example in Fig. 1. In the example, node $C$ has transmitted $P_i^j$ $(1 \leqslant i \leqslant 4, j = 1, 2)$ to receivers $R_j$ separately. As depicted in Fig 1(b), packets $P_1^1$, $P_1^2$, $P_2^1$ and $P_3^2$ have been failure to be received by their respective intended receiver but been correctly overheard by the other one. Traditionally, each one of $P_1^1$, $P_1^2$, $P_2^1$ and $P_3^2$ is retransmitted one by one and each one of them has only one intended receiver. With the NC-ARQ scheme, node $C$ can combine $P_1^1$ with $P_1^2$ to $P_{en}^{1,2} = P_1^1 \oplus P_1^2$, which is available for both $R_1$ and $R_2$. Similarly, we can do the same operation for $P_2^1$ and $P_3^2$. Even so, $P_4^2$ still has to be sent alone, since receiver $R_1$ doesn't obtain it, i.e. $R_1$ is unable to remove $P_4^2$ from the coded packet containing it to extract the intended packet. Furthermore, if $P_1^2$ and $P_2^1$ are lost again, node $C$ will XOR them together for next retransmission. From this example we can explicitly see that, by exploring the broadcast nature of wireless and mixing lost packets tog-ether, the total number of retransmissions can be effectively reduced.

Authors of [10] evaluated the performance of this method for single-hop scenarios, and got the following theorem.

**Theorem 2.** *Using the NC-ARQ technique, when the number of packets to be sent is sufficiently large, the average number of retransmissions $\lambda_{UNC}^N$ for the N-receiver unicast single-hop scenario is*

$$\lambda_{UNC}^N = \frac{1}{N} \sum_{i=1}^{N} \frac{\prod_{j=i}^{N} \omega_j}{1 - \omega_i} \qquad (7)$$

*where $\omega_j$ is the packet loss ratio between central node and receiver $R_j$, and $\omega_i \leqslant \omega_j$ if $1 \leqslant i \leqslant j \leqslant N$. In particular, for the 2-receiver scenario,*

$$\lambda_{UNC}^{2} = \frac{1}{2}\left(\frac{\omega_1\omega_2}{1-\omega_1} + \frac{\omega_2}{1-\omega_2}\right) \tag{8}$$

### *B. EAR performs on the N-reciever unicast scenario*

In this subsection, we evaluate the performance of the proposed EAR approach for the unicast topology. We first present how to use packet-loss pattern to identify the coding opportunities over retransmission process.

**Theorem 3.** *Let $R^{\rho_i}$ denote the set consisting of node(s) which is/are the intended receiver(s) for packets in $P_{\rho_i}$. In rescue process, the loss packets $P_1$, $P_2$,…,$P_n$ $(2 \leqslant n \leqslant N)$, each of which can be native or coded packet, with respect to the patterns $\rho_1, \ldots, \rho_n$ can be coded together if and only if* $R^{\rho_i} \cap R^{\rho_j} = \phi, i \neq j$ and *only the $i^{th}$ entry* of the $j_k^{th}$ *column of matrix $F = [\,\rho_1^T \ \ldots \ \rho_n^T\,]^T$ is equal to 0, where $R_{j_k} \in R^{\rho_i}$, $i \in [1, n]$* and $k \in [1, |\bigcup_{i=1}^{n} R^{\rho_i}|]$.

**Proof:** First of all, let us assume that there are two loss patterns $\rho_{i_1}$ and $\rho_{i_2} (i_1 \neq i_2, i_1, i_2 \in n)$ which satisfy $R^{\rho_{i_1}} \cap R^{\rho_{i_2}} = \{R^{cap}\}$. Suppose $P_{i_1}$ contains the native packet $P_1^{cap}$ and the corresponding packet in $P_{i_2}$ is $P_2^{cap}$. Under these assumptions, there would be two possibilities: 1. $P_1^{cap} \neq P_2^{cap}$, obviously, $R^{cap}$ cannot decode such coded packet; 2. $P_1^{cap} = P_2^{cap}$, then $R^{cap}$ cannot obtain this packet, for the reason $P_1^{cap} \oplus P_2^{cap} = 0$. Neither of them is expected by us. Hence, there must be $R^{\rho_i} \cap R^{\rho_j} = \phi, i \neq j$.

Without loss of generality, let us consider node $R_{j_1}$. To insure that the node $R_{j_1}$ get its corresponding packet $P_1$, it must have correctly overheard the packets $P_i, i = 2, \ldots, n$. So based on Definition 2, the $j_1^{th}$ entry in $\rho_i, i = 2, \ldots, n$ is equal to 1, i.e. the $j_1^{th}$ column of matrix *F* is only the first entry that is equal to 0. It is the same to other receivers. On the other hand, if the $j_k^{th}$ column *of* matrix *F* is only the $i^{th}$ entry that is equal to 0, where $R_{j_k} \in R^{\rho_i}$. It means that $R_{j_k}$ has already obtained the packets $P_t, t = 1, \ldots, n, t \neq i$, i.e. it enable to eliminate$P_t$ from the coded packet. Under the assumption, all packets except $P_i$ can be removed by $R_{j_k}$. Clearly, there are the same results for all other receivers. Therefore, based on the code rule of COPE [10], all the packets can be mixed together to one packet. ∎

**Definition 4.** *The set*

$$CG = \{\rho_1, \rho_2, \ldots, \rho_k\}, 2 \leqslant k \leqslant N \tag{9}$$

*where error patterns $\rho_1, \rho_2, \ldots, \rho_k$ can be coded together, is called the Code Group.*

Particularly, though there are no less than one code group interrelating with a error pattern, if all the $j_k^{th}$ $(j_k \notin \bigcup_{i=1}^{n} R^{\rho_i})$ columns of matrix $F$ must be zero in Theorem 5, and then each combination of loss patterns fulfilling the theorem 3 is unique, i.e if $\rho_i \in CG$ and $\rho_i \in CG'$, then $CG = CG'$, and achieve their maximum coding opportunities. In this way, all the code groups we mentioned are limited to this restriction in the rest of this paper.

In last sub-phase, we introduce the NC-ARQ approach. However, it only considered the native packets and ignored the encoded packets, which contain coding opportunities and have potential to enhance the retransmission efficiency as well. We use induction method to prove the lemma. Interested readers can find details of the proof in the Appendix.

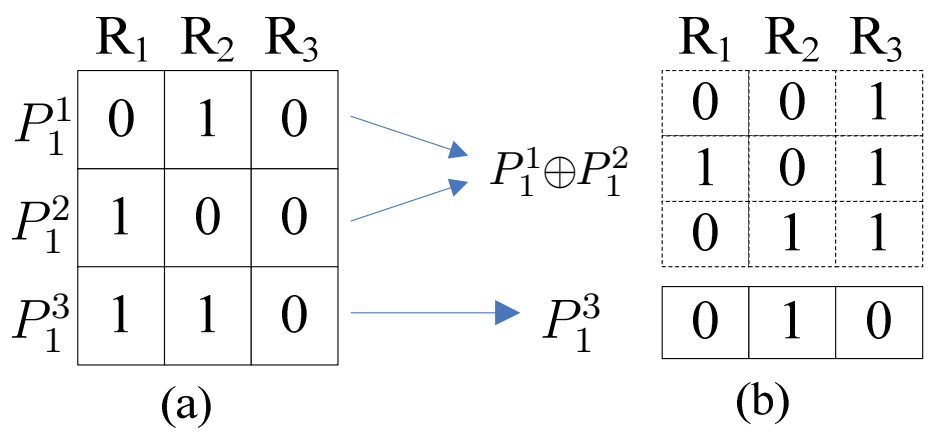

Figure 2. Three receivers uncast scenario.

Let's take an example in Fig. 2 to illustrate how to utilize them to improve the retransmission efficiency.

In this example, node *C* has transmitted $P_1^j$ $(j = 1, 2, 3)$ to receivers $R_j$ separately. The receive-state of packets $P_1^1$, $P_1^2$ and $P_1^3$ is depicted in Fig 6(a). Due to the assumption, only $P_1^1$ and $P_1^2$ can be XORed together. Therefore, node *C* mixes them together and forwards encoded packet $P_1^1 \oplus P_1^2$ to the downstream nodes. However, $P_1^1 \oplus P_1^2$ would be lost again over the noisy channel, if its receive-state is the showed in Fig. 6(b), then the best coding decision would be to combine $P_1^1 \oplus P_1^2$ and $P_1^3$ together, and not to deliver them respectively. From this example we can explicitly see that, by considering the encoded packets generated in rescue process, the number of retransmissions would be effectively reduced.

**Lemma 1.** *Using the NC-ARQ technique, for N-receiver scenario, if $\rho_i^{r_1}$ and $\rho_j^{r_2}$ can be coded together and $|P_{\rho_i}^{r_1}| \leqslant |P_{\rho_j}^{r_2}|$, $\omega_{r_1} \leqslant \omega_{r_2}$, then the entire packets in $P_{\rho_i}^{r_1}$ can be XORed with the ones belonging to $P_{\rho_j}^{r_2}$ in recovery process, and we say that $\rho_i^{r_1}$ is dominated by $\rho_j^{r_2}$. This relation is denoted by $\rho_i^{r_1} \succ \rho_j^{r_2}$. In particular, node C need to retransmit $\Omega_{\rho_i}^{r_1}$ times to rescue $\rho_i^{r_1}$. And $\Omega_{Native}$ native packets would be sent alone.*

$$\Omega_{\rho_i}^{r_1} = \frac{|P_{\rho_i}^{r_1}|}{1-\omega_{r_1}} \tag{10}$$

$$\Omega_{Native} = |P_{\rho_j}^{r_2}| - \frac{|P_{\rho_i}^{r_1}|(1-\omega_{r_2})}{1-\omega_{r_1}} \tag{11}$$

**Theorem 4.** *Using the EAR method, when the number of packets to be sent is sufficiently large, the average number of retransmissions $\lambda_{UNC}^{E}$ for the N-receiver unicast single-hop scenario is*

$$\lambda_{UNC}^{E} = \frac{1}{N}\sum_{i=1}^{N}\frac{\prod_{j=i}^{N}\omega_j}{1-\prod_{j=i}^{N}\omega_j} \tag{12}$$

*where $\omega_j$ is the packet loss ratio between central node and receiver $R_j$, and $\omega_i \leqslant \omega_j$ if $1 \leqslant i \leqslant j \leqslant N$.*

**Proof:** First, we consider the 2-receiver scenario. Let $\rho_0^1 = [0 \ \ 0]$, $\rho_1^1 = [0 \ \ 1]$, $\rho_0^2 = [0 \ \ 0]$ and $\rho_1^1 = [1 \ \ 0]$. Let random variable $X_0$, $X_1$, $X_2$ and $X_3$, separately, denote the number of lost packets relevant to loss pattern $\rho_0^1$, $\rho_1^1$, $\rho_0^2$ and $\rho_1^2$ after *K* transmissions. $X_k \quad (k = 0, 1, 2, 3)$ follow the binomial distribution. As we mentioned above, node *C* would rescue $\rho_0^1$ and $\rho_0^2$ first. Then based on Theorem 1, we have

$$\Omega_{\rho_0^1} = \Omega_{\rho_0^2} = \frac{K\omega_1\omega_2}{1-\omega_1\omega_2} \quad (13)$$

$$\Omega_{\rho_0^1 \to \rho_1^1} = \Omega_{\rho_0^1} \cdot \omega_1(1-\omega_2) \quad (14)$$

$$\Omega_{\rho_0^1 \to \rho_1^2} = \Omega_{\rho_0^2} \cdot (1-\omega_1)\omega_2 \quad (15)$$

Since $\Omega_{\rho_0^1 \to \rho_1^1} \leqslant \Omega_{\rho_0^2 \to \rho_1^2}$ and $E[X_1] \leqslant E[X_3]$, then $E[X_1] + \Omega_{\rho_0^1 \to \rho_1^1} = m_1 \leqslant m_2 = E[X_2] + \Omega_{\rho_0^2 \to \rho_1^2}$. Based on Lemma 1, we directly get $\rho_1^1 \succ \rho_1^2$, and

$$\Omega_{\rho_1^2} = \frac{m_2}{1-\omega_2} \quad (16)$$

Combining (13) and (16), the expected number of retransmissions to successfully deliver $K$ encoded packets to $R1$ and $R2$ is given by

$$\Omega_{all} = \Omega_{\rho_0^1} + \Omega_{\rho_0^2} + \Omega_{\rho_1^2} = \frac{K\omega_2}{1-\omega_2} + \frac{K\omega_1\omega_2}{1-\omega_1\omega_2} \quad (17)$$

The proof for $N$(>2) receivers scenario will be found in the Appendix. ■

**Remark:** Though no encoded packets would be utilized in 2-receiver unicast scenario, the EAR approach also outperforms the NC-ARQ technique as shown in Fig. 3, where $\omega_1 = \omega_2$. The reason is that the EAR approach not only studies the potential of encoded packets, but also considers that loss packets should synchronously update their loss-state over rescue process, which would result in more coding opportunities.

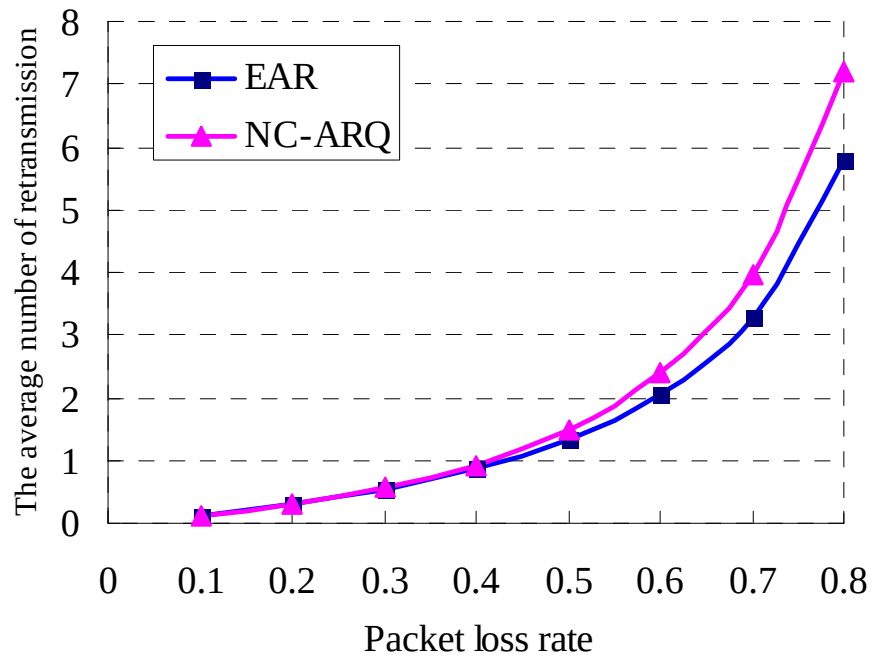

Figure 3. The average number of retransmissions on 2-receiver scenario.

Note that, in proof we suppose each coded packet can be divided into several native packets, all of which are the same receive-state and would be recovered separately, even the receivers cannot decode it. However, this hypothesis does clearly not hold in actual applications. As a result, some encoded packets are unable to be mixed with any other packet. We call them the *unwanted packets*. We still take the 3-receiver unicast scenario to illustrate how it happens.

The zero weight patterns and $CG = \{\rho_3^1, \rho_3^2, \rho_3^3\}$ don't produce undecodabe packets in each node. Consequently, there are only three coding combinations that would create unwanted packets, meanwhile, though each coded packet has six possible loss-state, only three of them as shown in Fig. 4(a) would generate undecodable packets in unintended node. Based on Theorem 3, the potential coding schemes for these patterns are depicted in Fig. 4(b).

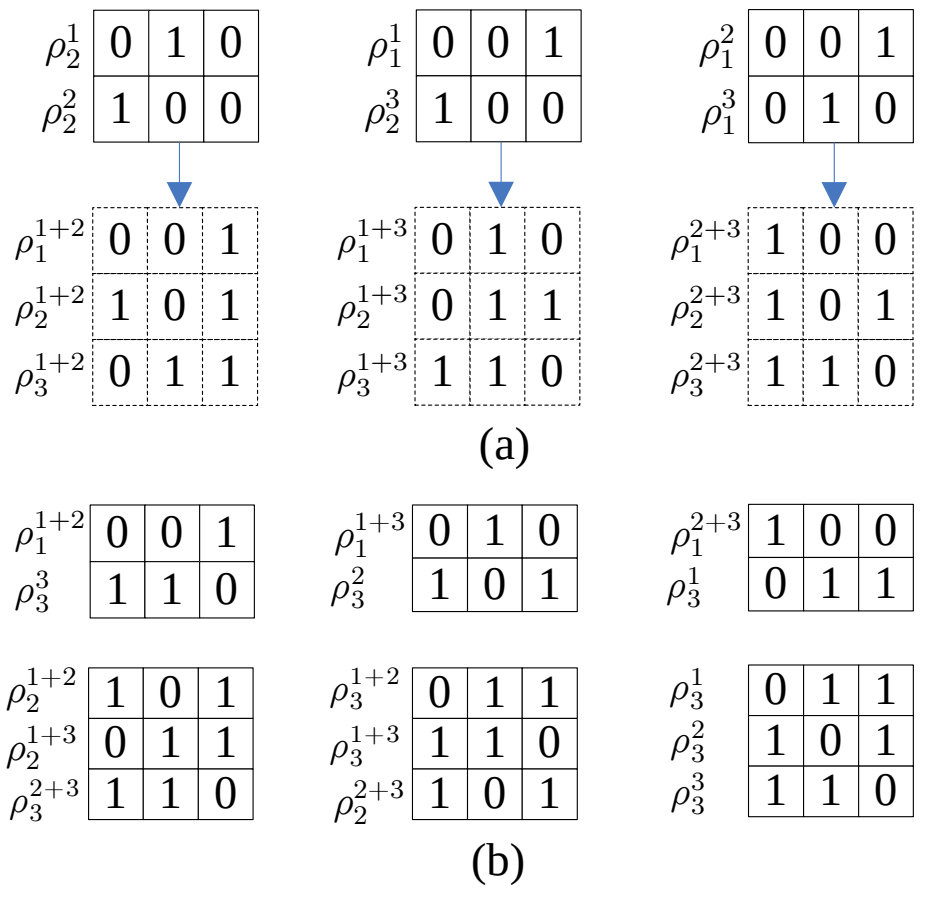

Figure 4. Example of the influence of unwanted packets. (a) The effective coded patterns. (b) The coding schemes for coded patterns.

To simplify analysis, we assume $\omega_1 = \omega_2 = \omega_3 = \omega$. In the light of Theorem 1, we have

$$N_1 = |P_{\rho_3}^1| = |P_{\rho_3}^2| = |P_{\rho_3}^3| = \frac{K\omega(1-\omega)^2}{1-\omega^3}$$

$$|P_{\rho_1}^1| = |P_{\rho_2}^1| = |P_{\rho_1}^2| = |P_{\rho_2}^2| = |P_{\rho_1}^3| = |P_{\rho_2}^3|$$

and we carry out

$$|P_{\rho_2}^{1+2}| = |P_{\rho_3}^{1+2}| = |P_{\rho_2}^{1+3}| = |P_{\rho_3}^{1+3}| = |P_{\rho_2}^{2+3}| = |P_{\rho_3}^{2+3}|$$

$$N_2 = |P_{\rho_1}^{1+2}| = |P_{\rho_1}^{1+3}| = |P_{\rho_1}^{2+3}| = \frac{K\omega^4(1-\omega)^2}{(1-\omega^3)(1-\omega^2)}$$

Clearly, $CG_1 = \{\rho_2^{1+2}, \rho_2^{1+3}, \rho_3^{2+3}\}$ and $CG_2 = \{\rho_3^{1+2}, \rho_3^{1+3}, \rho_2^{2+3}\}$ would not create unwanted packets, for the reason that the expect number of retransmissions requested to rescue each error pattern is the same. Moreover, if $N_1 > N_2$, then $\rho_1^{1+2} \succ \rho_3^3$, $\rho_1^{1+3} \succ \rho_3^2$ and $\rho_1^{2+3} \succ \rho_3^1$, and the remaining packets in $P_{\rho_3}^3$, $P_{\rho_3}^2$ and $P_{\rho_3}^1$, respectively, are the same size based on Lemma 1. Since $\rho_3^1$, $\rho_3^2$ and $\rho_3^3$ can be XORed together, $CG_3 = \{\rho_1^{1+2}, \rho_3^3\}$, $CG_4 = \{\rho_1^{1+3}, \rho_3^2\}$ and $CG_5 = \{\rho_1^{2+3}, \rho_3^1\}$ would not produce unwanted packets either. On the contrary, if $N_1 < N_2$, then the residual packets in $P_{\rho_1}^{1+2}$, $P_{\rho_1}^{1+3}$ and $P_{\rho_1}^{2+3}$ have to be recovered separately. However, the packets in $P_{\rho_1}^{1+3}$ and $P_{\rho_1}^{2+3}$ are not unwanted packets, in respect that they still combine with the packets requested by node $R_3$. Therefore, only the one in $P_{\rho_1}^{1+2}$ is called unwanted packet. The expect number of deliveries to recover them is

$$\Omega = \frac{K\omega(1-\omega)(\omega^3+\omega^2-1)}{(1-\omega^3)(1-\omega^2)} \quad (18)$$

where $\omega^3 + \omega^2 - 1 > 0$.

Though Theorem 4 only provides the upper bound of the performance of the EAR approach on N-receiver single-hop unicast scenario, the simulation results indicate that the influence of the unwanted packets is small and the performance degradation it caused is considered negligible.

```
Updating algorithm:
ρ_P = 0
Triumph = 0
Transform = false
for receivers i = 1 to N do
  if R_i has obtained P correctly then
    if R_i is one of the destination of P then
      Triumph = Triumph + 1
      if Triumph is equal to NC(P) then
        delete P from the retransmission queue P
        if P is a coded packet then
          delete the corresponding native packets contained in P
        end if
        return
      end if
      set the i^th entry of ρ_P equal to 1
    else
      if R_i has gain P for the first time then
        set the i^th entry of ρ_P equal to 1
      end if
      Transform = true
    end if
  end if
end for
if Transform is true and P is a coded packet then
  delete the corresponding native packets contained in P
end if
```

Figure 5. Updating algorithm for each loss packet

Eventually, node $C$ must update the receive-state of loss packets synchronously to effectively implement EAR. Thus, after $C$ has received $N$ feedback from $R_i$, $i = 1, \ldots, N$ for a packet $P$, it will execute the following updating algorithm, where $NC(P)$ denotes the number of the native packets contained in $P$. Notice that when a coded packet is lost by its intended receivers and its loss-state is changed, then node $C$ would only reserve these packets and discard the corresponding native packets comprising it to save memory.

## IV. Experimental Results and Discussions

In this section, retransmission gain is used to evaluate the retransmission efficiency of different approaches by varying the number of receivers and bit error rate (BER) under both unicast and wheel topology. The BER at each receiver are mutually independent and follow the Bernoulli distribution. We define the retransmission gain as the total number of retransmissions using a typical retransmission algorithm, which is the HARQ or NC-HARQ techniques, divided by the total number of retransmissions using our algorithm. A higher retransmission gain is preferred since it indicates fewer retransmissions. In the following simulate, the packet size is to be 1532 bytes and data is encoded with RS (32, 28, 4). We use CRC-16 for error detection in all the simulations. We record the total number of retransmissions over a mass of experiments. In the interest of space and clarity, we only present the average retransmission gains in the following simulation results.

First, we compare the performance of proposed EAR approach with the NC-HARQ technique and the HARQ technique. Fig. 6 shows the effect of different BERs on the retransmission gain for the 3-receiver unicast scenarios, respectively. BERs between the sender and all its receivers are the same, and varied from $10^{-4}$ to $3.5 \times 10^{-3}$ in $5 \times 10^{-4}$ increments. As expected, the simulation results support our theoretical derivations. Though in large BERs region the unwanted packets would bring down the performance of the EAR approach as we discussed before, the simulation results for unicast scenario show such decline is so small as to be neglectable. Furthermore, an interesting phenomenon would be observed is that the retransmission gains for unicast scenario increase with BER, and are very close to 1 under the light BERs. The reason is that the more awful noise the channels encounter, the more packets are lost, that is to say the more encoded packets would be generated from which the EAR protocol enables to explore the coding opportunities, yet NC-ARQ cannot.

Then, we compare the performance of proposed EAR approach with the NC-HARQ technique versus the number of clients for unicast and wheel topology. In the experiment, BERs between the sender and all its receivers are set equal to $2 \times 10^{-3}$ and $3 \times 10^{-3}$. Fig. 7 shows the impact of different number of receivers, which is varied from 3 to 25, on the retransmission gain. As seen, the retransmission gains hardly increase with the number of clients on the light BER, whereas the gains significantly advanced upon the heavy BER with the slowing growth rate. The reason is that the probability that a loss packet is successfully received at lease at one client increases with the number of the clients, and then more loss packets are relate to one node, namely the growth rate of the total number of retransmissions is getting slower and slower, when the number of receivers goes up.

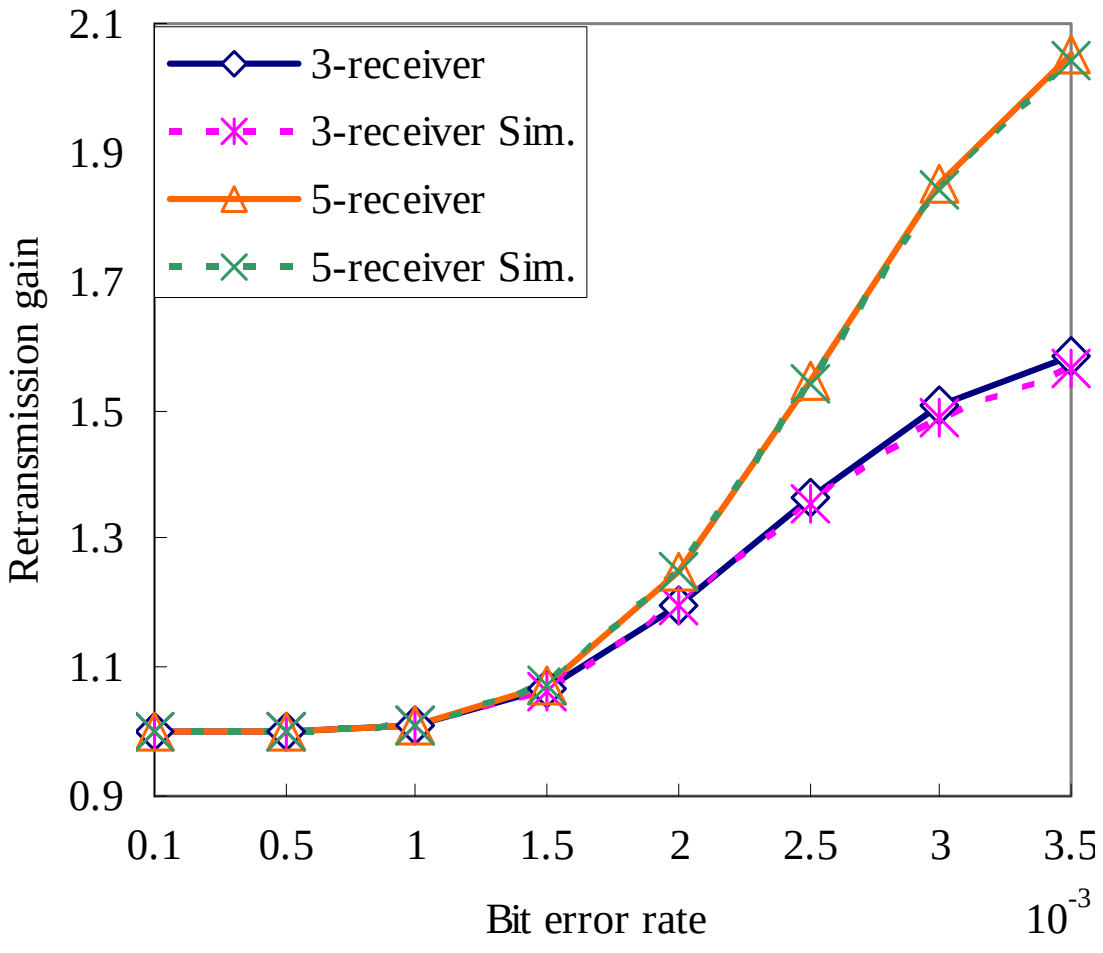

Figure 6. retransmission gain versus BER for theory and simulation.

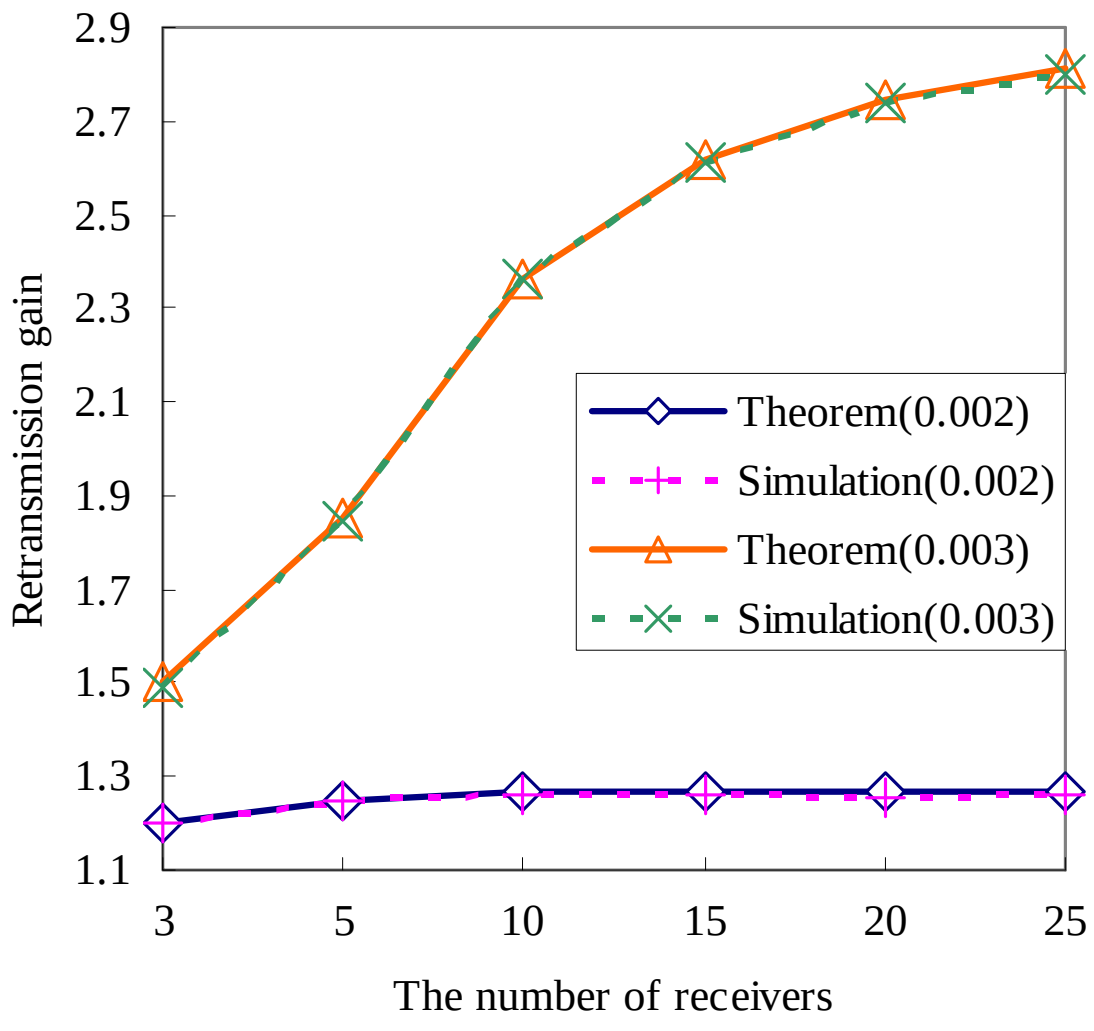

Figure 7. Retransmission gain versus the number of receivers for theory and simulation.

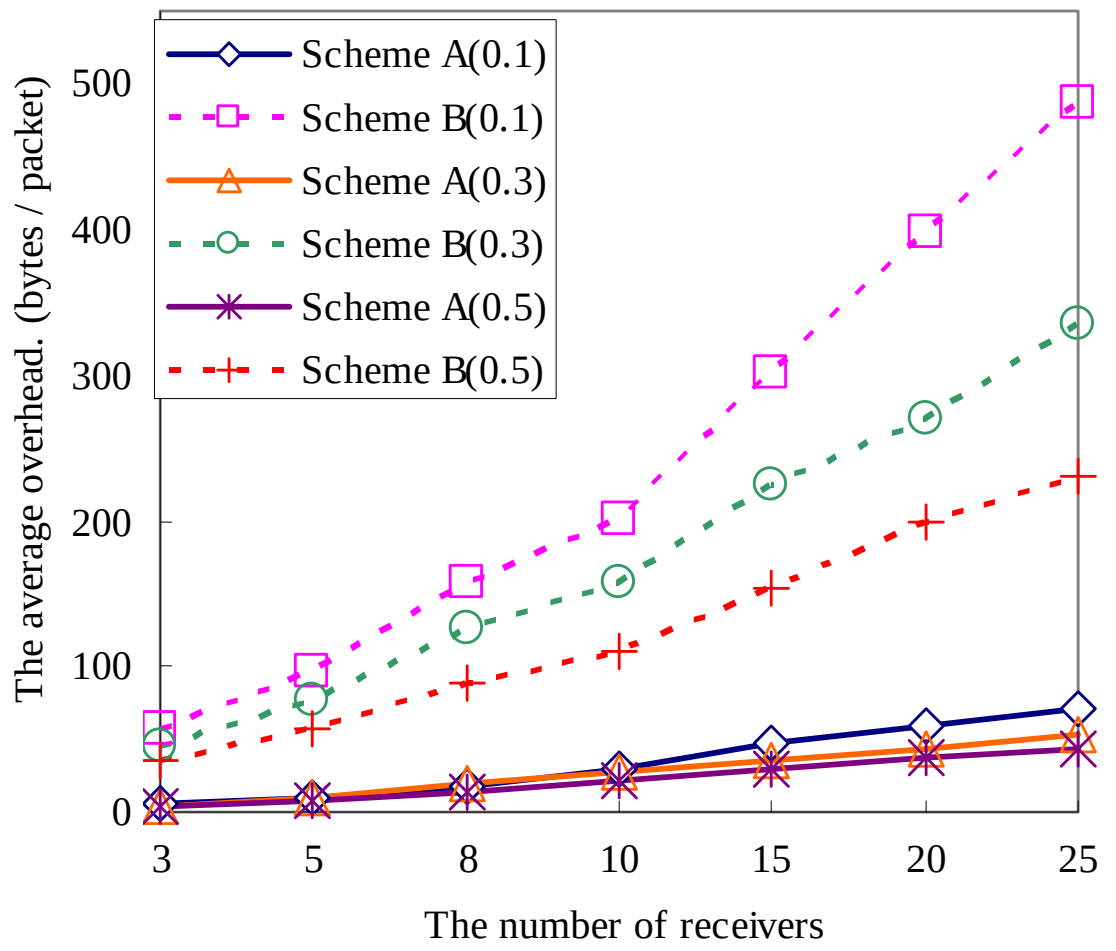

Figure 8. The average overhead

Finally, to ensure each intended receiver is able to successfully decode a coded packet, node $C$ should add the information, which specifies whether it consists of encoded packets and which encoded packets are embraced, within the NC header. To achieve this goal, there are two schemes.

**Scheme A**: node $C$ directly records all the native packets that construct the coded packets, regardless whether any one of them is or was a loss packet. We utilize a 2-byte hashvalue to identify a packet's source address and sequence number. Thereupon, in the worst condition, node $C$ has to use $2 \times K \times (\omega_1 + \cdots + \omega_N)$ bytes to record these packets.

**Scheme B**: node $C$ employs "bit map" to register these native packets. In this scenario, $C$ has to keep 19-byte for each destination, owing to the randomicity of loss-state of each packet. Hence, in the worst condition, node $C$ needs to append $19 \times N$ bytes.

Fig. 8 shows the average length of header that records which packets in the combined packets for both schemes. Packet loss rate between the sender and all its receivers are the same and equal to 0.1, 0.3 and 0.5, respectively. Through the figure, we see that scheme A has much smaller overhead than scheme B. And we further notice that the overhead increases with the number of receivers, but decreases when BERs rises. Even so, comparing to the length of the data packets, the overhead associated with scheme A is no more than 5%, which can be neglected.

## V. CONCLUTIONS

In this paper, we addressed the problem of existing NC-based reliable transmission schemes for wireless multi-unicast system such as WiFi and WiMAX networks. In order to overcome the disregard for the role of encoded packets, an encoded packet-assisted rescue approach (EAR) based on XOR network coding has been proposed. The core features of the EAR approach are 1) explore the coding opportunities of encoded packet; 2) synchronously updating the receive-state of loss packets. Under the access point model, some analytical results are derived for the retransmission effi-ciency over the unicast scenarios. The theoretical and simu-lation results show that the proposed EAR approach always outperforms the NC-HARQ technique in terms of the retran-smission efficiency for a typical range of wireless network conditions.

## REFERENCES

[1] D. Aguayo, J. Bicket, S. Biswas, G. Judd, and R. Morris. "Link-level Measurements from an 802.11b Mesh Network," in Proc. of ACM SIGCOMM 04, August 2004, pp. 121-132.

[2] M. Naijoh, S. Sampei, N. Morinaga, and Y. Kamio, "ARQ schemes with adaptive modulation/TDMA/TDD systems for wireless multi-media communication services," The 8th IEEE International Symp. on PIMRC '97, Sept. 1997, pp. 709–713.

[3] D. J. Costello, J. Hagenauer, H. Imai, and S. B. Wicker, "Application of error-control coding," IEEE Transactions on Information Theory, vol. 44, Oct. 1998, pp. 2531–2560.

[4] Jung-Fu Cheng, "Coding performance of hybrid ARQ schemes," IEEE Trans. on Conmmun., vol. 54, no.6, June 2006, pp. 1017–1029.

[5] R. Ahlswede, N. Cai, S.-Y. R. Li, and R. W. Yeung, "Network information flow," IEEE Transactions on Information Theory, vol.46, no.2, July 2000, pp. 1204–1216.

[6] S. Katti, H. Rahul, W. Hu, D. Katabi, M. Medard, and J. Crowcroft, "XORs in The Air: Practical Wireless Network Coding," in Proc. of ACM SIGCOMM'06, July 2006.

[7] S. Sengupta, S. Rayanchu and S. Banerjee, "An Analysis of Wireless Network Coding for Unicast Sessions: The Case for Coding-Aware Routing", in Proc. of IEEE INFOCOM'07, May 2007, pp. 1026-1038.

[8] D. Nguyen, T. Nguyen and B. Bose, "Wireless broadcasting using network coding," in Third Workshop on Network Coding, Theory, and Applications, Jan. 2007.

[9] E. Rozner, A. P. Iyer, Y. Mehta, L. Qiu, and M. Jafry, "ER: efficient retransmission scheme for wireless lans," in Proc. of the 2007 ACM CoNEXT conference, Dec. 2007, pp. 1-12.

[10] T. Tran, T. Nguyen, B. Bose, and V. Gopal "A Hybrid Network Coding Technique for Single-Hop Wireless Networks," IEEE Journal on Selected Areas in Com., vol. 27, no. 5, June 2009, pp. 685-698..

[11] Z. Zheng, and P. Sinha, "XBC: XOR-based buffer coding for reliable transmissions over wireless networks," in: Proc. IEEE BROADNETS, Sep. 2007, pp. 76-85.

[12] M. Ghaderi, D. Towsley, and J. Kurose, "Reliability Gain of Network Coding in Lossy Wireless Networks", in Proc. of IEEE INFOCOM, April 2008, pp. 2171 - 2179.

[13] A. A. Yazdi, S. Sorour, S. Valaee, and R. Y. Kim "Optimum Network Coding for Delay Sensitive Applications in WiMAX Unicast," in Proc. of INFOCOM'09, April 2009, pp. 2576-2580.

[14] H. Sofiane, M. Patrick, and R. David, "On the Impact of Random Losses on TCP Performance in Coded Wireless Mesh Networks," in Proc. of IEEE INFOCOM'10, March 2010.

## APPENDIX

**Proof of lemma 1:** We still suppose that the central node retransmits $P_{\rho_i}^{r_1}$ round by round. We define $P_{\rho_i}^{r_1}(k)$ and $P_{\rho_j}^{r_2}(k)$ $(k > 0)$ as the loss packets sets relevant to $\rho_i^{r_1}$ and $\rho_j^{r_2}$ after the $k^{th}$ round and we get $P_{\rho_i}^{r_1}(0) = P_{\rho_i}^{r_1}$, $P_{\rho_j}^{r_2}(0) = P_{\rho_j}^{r_2}$. Let random variable $Y_k$ and $Z_k(k > 0)$ represent the cardinality of $P_{\rho_i}^{r_1}(k)$ and $P_{\rho_j}^{r_2}(k)$, furthermore, we define $Y_0 = |P_{\rho_i}^{r_1}|$, $Z_0 = |P^{r_2}_{\rho_j}|$. As the deliveries are i.i.d. and follow the Bernoulli distribution, random variables $Y_k$ and $Z_k$ $(k = 0, 1, \dots)$ are i.i.d too and follow the Binomial distribution. Due to $Y_0 \leqslant Z_0$, the entire packets in $P^{r_1}_{\rho_i}(0)$ is able to be combined with the ones belonging to $P^{r_2}_{\rho_j}(0)$ in next round. After the first round, $Y_1 = \omega_{r_1} Y_0$ and $m_0 = Z_0 - Y_0$ packets in $P^{r_2}_{\rho_j}(0)$ are not delivered, i.e. $Z_1 = \omega_{r_2} Y_0 + m_0$. Because of $\omega_{r_1} \leqslant \omega_{r_2}$, we explicitly have $Y_1 \leqslant Z_1$. It means that the entire packets within $P^{r_1}_{\rho_i}(1)$ can be mixed with the ones belonging to $P^{r_2}_{\rho_j}(1)$ over the next round in the same way.

Now, we suppose that $Y_n \leqslant Z_n$ $(n \geqslant 3)$. Then after the $n^{th}$ round, node $R_1$ is failure to receive $Y_{n+1} = \omega_{r_1} Y_n$ packets, so the central node has to retransmit these packets in the next round. Moreover, $m_n = Z_n - Y_n$ packets in $P^{r_2}_{\rho_j}(n)$ are not delivered, i.e. $Z_{n+1} = \omega_{r_2} Y_n + m_n$. Clearly, we have $Y_{n+1} \leqslant Z_{n+1}$, which indicates that we are able to combine the entire packets in $P^{r_1}_{\rho_i}(n+1)$ with the ones belonging to $P^{r_2}_{\rho_j}(n+1)$. Then we deduce $\rho_1 \succ \rho_2$ by induction, and work out the expect number of retransmissions required to rescue $\rho_1$.

$$\Omega'_{Code} = \sum\nolimits_{k=0}^{\infty} Y_k = \frac{|P_{\rho_1}|}{1-\omega_{r_1}} \tag{19}$$

In particular, $\Omega'_{Native}$ native packets relevant to $\rho_2$ would be delivered alone.

$$\begin{aligned} \Omega'_{Native} &= Z_\infty \\ &= Z_0 + \sum\nolimits_{k=0}^{\infty} (\omega_{r_2} - 1) Y_k \\ &= |P_{\rho_2}| - \frac{|P_{\rho_1}|\,(1-\omega_{r_2})}{1-\omega_{r_1}} \end{aligned} \tag{20}$$

The lemma has been proved. ∎

**Proof of theorem 7:** First, we separate each coded pattern into the corresponding native patterns, in respect that a coded pattern can be considered as the set consisting of loss packet(s) requested by distinct node(s) with the same error-state. For example, if a encoded packet $P$ in $P^{r_1,r_2,r_3}_{\rho_j}$ contains loss packets $P^{r_1}_{i_1}$ and $P^{r_2}_{i_2}$, then we put $P^{r_1}_{i_1}$ and $P^{r_2}_{i_2}$ into $P^{r_1}_{\rho_{j'}}$ and $P^{r_2}_{\rho_{j''}}$ respectively, where $\rho_j = \rho_{j'} = \rho_{j''}$. After that, we cancel $P$ from $P^{r_1,r_2,r_3}_{\rho_j}$ and repeat these steps until $P^{r_1,r_2,r_3}_{\rho_j} = \phi$. In this way, we only consider native packets in the following proof.

Second, let $\Phi_i$ denote the set that contains the error patterns which belong to $\rho^j$ $(j < i)$ and the $i^{th}$ entry is non-zero, particularly $\Phi_0 = \phi$. Let $\Phi^i$ denote the subset of $\rho^i$ consisting of all those error patterns whose the $j^{th}$ $(j \geqslant i)$ entries are zero, specially $\Phi^0 = \{\rho^0_0\}$, where $\rho^0_0$ is zero vector, and $\Phi^N = \rho^N$. Then the set $U^i = \Phi_i \cup \Phi^i$ is called the primary set for node $R_i$,. Clearly, $U^i \cap U^j = \phi$, $i \neq j$ and $\bigcup_{i=1}^N U^i = \bigcup_{i=1}^N \rho^i$.

Based on the above concepts and Theorem 5, for any loss pattern $\rho^i_k$ in $\Phi^i$, there is a unique code group, $CG = \{\rho^i_k, \rho^{i_1}_{k_1}, \dots, \rho^{i_w}_{k_w}\} \subseteq U^i$, where $w = W(\rho^i_k)$, $i_j \in N$ and $k_j \in \left[0, \left|\rho^{i_j}_{k_j}\right|\right]$. Furthermore, only the $i^{th}$ and $i^{th}_j$ entries of vector $\rho^i_k \oplus \rho^{i_j}_{k_j}$ are equal to 1, and, $\omega_{i_j} \leqslant \omega_i$, thereby $\Pr\{\rho^i_k\} \geqslant \Pr\{\rho^{i_j}_{k_j}\}$. We define the loss packets set $P^{(n)}_\rho$ relating to $\rho \in CG$ after the $n^{th}$ round, and $P^{(0)}_\rho = P_\rho$. Let random variable $Y_n = |P^{(n)}_\rho|$ $(n > 0)$. Since the deliveries are i.i.d. and follow the Bernoulli distribution, random variables $Y_n$ $(n = 0, 1, \dots)$ are i.i.d and follow the Binomial distribution. Then using the EAR approach, when the intended node(s) is/are failure to receive a encoded packet and some client overheard it for the first time, these contained native packets would be eliminated from their original sets and transferred to the matched sets respectively. Therefore, after the $n^{th}$ round, $Y_{n+1} = Y_n \Pr\{\rho|\rho\}$ packets have to be retransmitted in the next round. Notice that it differs from the result in Lemma 1. This is due to the fact that the NC-ARQ technique is only interest in whether the intended nodes have already correctly obtains the coded packet, yet the EAR approach not merely takes care of it, but also considers how to combine more requested packets together whether they are encoded packets or not. Then to rescue $\rho$, node $C$ is expected to retransmit $\Omega_\rho$ times,

$$\Omega_\rho = \sum\nolimits_{k=0}^{\infty} Y_n = \frac{K|P_\rho|}{1 - \Pr\{\rho|\rho\}} \tag{21}$$

and $\Omega_{\rho \to \rho'}$ packets is transformed from $P_\rho$ to $P_{\rho'}$.

$$\Omega_{\rho \to \rho'} = \Omega \times \Pr\{\rho'|\rho\} \tag{22}$$

Note that this result is the same as Theorem 1.

On the other hand, some loss packets relating to $\rho' \in U^t$ $(t \leqslant i)$ would be relocated to $P_\rho$ $(\rho \in CG)$, if $\Pr\{\rho|\rho'\} > 0$. And based on the definition of the sets $\Phi^i$ and $\Phi_i$, there is always an error pattern $\rho^i_x \in \Phi^i = \rho^{i_j}_{x'} \in U^t (t < i)$ and $\rho^i_y \in \Phi^i$ enables to be combined with $\rho^{i_j}_{y'} \in \Phi_i$. Because the $i^{th}_j$ entry of $\rho^i_x$ is zero and of $\rho^i_y$ is non-zero, then $\{\rho^i_x\} \cap \{\rho^i_y\} = \phi$ and $\{\rho^i_x\} \cup \{\rho^i_y\} = \Phi^i$. Due to $\rho^i_x = \rho^{i_j}_{x'}$, then we have $\Pr\{\rho^i_x\} = \Pr\{\rho^{i_j}_{x'}\}$, $\Pr\{\rho^i_x|\rho^i_z\} = \Pr\{\rho^{i_j}_{x'}|\rho^{i_j}_{z'}\}$ ( $\rho^i_z = \rho^{i_j}_{z'}$ ), $|P^i_{\rho_x}| = |P^{i_j}_{\rho_{x'}}|$ and $\Pr\{\rho^i_k|\rho^i_x\} \geqslant \Pr\{\rho^{i_j}_{k_j}|\rho^{i_j}_{x'}\}$, namely $\Omega_{\rho^i_x} = \Omega_{\rho^{i_j}_{x'}}$ and $\Omega_{\rho^i_x \to \rho^i_k} \geqslant \Omega_{\rho^{i_j}_{x'} \to \rho^{i_j}_{k_j}}$. In addition, we also have

$$\begin{aligned} \Pr\{\rho^i_y\} &= \Pr\{\rho^{i_j}_{y'}\} \\ \Pr\{\rho^i_k|\rho^i_y\} &= \Pr\{\rho^{i_j}_{k_j}|\rho^{i_j}_{y'}\} \end{aligned} \tag{23}$$

for the reason $\rho^i_k \oplus \rho^{i_j}_{k_j} = \rho^i_y \oplus \rho^{i_j}_{y'}$.

Hence, we divide the loss packet set into three parts

$$\begin{aligned} P^i_{\rho_k} &= Or^i_{\rho_k} \cup \{\textstyle\bigcup_{\rho^i_x} (\Omega_{\rho^i_x \to \rho^i_k})\} \cup \{\textstyle\bigcup_{\rho^i_y} (\Omega_{\rho^i_y \to \rho^i_k})\} \\ P^{i_j}_{\rho_{k_j}} &= Or^{i_j}_{\rho_{k_j}} \cup \{\textstyle\bigcup_{\rho^{i_j}_{x'}} (\Omega_{\rho^{i_j}_{x'} \to \rho^{i_j}_{k_j}})\} \cup \{\textstyle\bigcup_{\rho^{i_j}_{y'}} (\Omega_{\rho^{i_j}_{y'} \to \rho^{i_j}_{k_j}})\} \end{aligned}$$

Due to $|Or^i_{\rho_k}| = K \Pr\{\rho^i_k\} \geqslant |Or^{i_j}_{\rho_{k_j}}| = K \Pr\{\rho^{i_j}_{k_j}\}$, we have

$$\left|Or^i_{\rho_k} \cup \{\textstyle\bigcup_{\rho^i_x} (\Omega_{\rho^i_x \to \rho^i_k})\}\right| \geqslant \left|Or^{i_j}_{\rho_{k_j}} \cup \{\textstyle\bigcup_{\rho^{i_j}_{x'}} (\Omega_{\rho^{i_j}_{x'} \to \rho^{i_j}_{k_j}})\}\right|.$$

Now, we assume $W(\rho^i_k) = 1$, then $\{\bigcup_{\rho^{i_j}_{y'}} (\Omega_{\rho^{i_j}_{y'} \to \rho^{i_j}_{k_j}})\} =$

$\{\bigcup_{\rho_y^i}(\Omega_{\rho_y^i\to\rho_k^i})\}=\phi$, thereby $|P_{\rho_k}^i|=\max_{\rho\in CG}\{|P_\rho|\}$. If $W(\rho_k^i) = 2$, then $|\Omega_{\rho_y^i\to\rho_k^i}|\neq 0$ unless $W(\rho_y^i)=1$. Based on (21), (22) and (23), we get $|\Omega_{\rho_y^i\to\rho_k^i}|\geqslant|\Omega_{\rho_{y'}^{i_j}\to\rho_{k_j}^{i_j}}|$ for each pair $(\rho_y^i,\rho_{y'}^{i_j})$, so $|P_{\rho_k}^i|=\max_{\rho\in CG}\{|P_\rho|\}$. Suppose $|P_{\rho_k}^i|=\max_{\rho\in CG}\{|P_\rho|\}$ is valid if $W(\rho_k^i)=n-1$ $(n<i)$. Then for $W(\rho_k^i)=n$, there are

$$\{\bigcup_{\rho_x^i}(\Omega_{\rho_x^i\to\rho_k^i})\}=\{\bigcup_{w=1}^{n-1}(\Omega_{\rho_x^i\to\rho_k^i}(w))\}$$

$$\{\bigcup_{\rho_{y'}^{i_j}}(\Omega_{\rho_{y'}^{i_j}\to\rho_{k_j}^{i_j}})\}=\{\bigcup_{w=1}^{n-1}(\Omega_{\rho_{y'}^{i_j}\to\rho_{k_j}^{i_j}}(w))\}$$

where $\Omega_{\rho\to\rho'}(w)$ is the set consisting of all the $\Omega_{\rho\to\rho'}$ ($W(\rho)=w$). Because of the assumption, for each pair ($\rho_y^i$, $\rho_{y'}^{i_j}$), there are $W(\rho_y^i)<n$, and we have $|P_{\rho_y}^i|\geqslant|P_{\rho_{y'}}^{i_j}|$, then according to (21), (22) and (23), we still get $|\Omega_{\rho_y^i\to\rho_k^i}|\geqslant|\Omega_{\rho_{y'}^{i_j}\to\rho_{k_j}^{i_j}}|$. It is able to deduce $\Omega_{\rho_x^i\to\rho_k^i}(w)\geqslant\Omega_{\rho_{y'}^{i_j}\to\rho_{k_j}^{i_j}}(w)$, $w\in[1,n-1]$. Therefore, $|P_{\rho_k}^i|=\max_{\rho\in CG}\{|P_\rho|\}$.

Then by induction, the equation $|P_{\rho_k}^i|=\max_{\rho\in CG}\{|P_\rho|\}$ is available at all time for each $\rho_k^i\in\Phi^i$, where *CG* contains $\rho_k^i$. Based on (21), we have $\Omega_{\rho_k^i}\geqslant\Omega_{\rho_{k_j}^{i_j}}$, namely after the central node retransmits $\Omega_{\rho_k^i}$ times, *CG* has been already rescued, in other words *CG* is dominated by $\rho_k^i$. Consequently, $U^i\prec\Phi^i$ and to rescue $U^i$ is equivalent to recover loss packets in $P_\rho$, $\rho\in\Phi^i$.

Furthermore, we define that when the loss-state of a packet in $Or_\rho$ ($\rho\in\Phi^i$) is altered, but still belongs to $\Phi^i$, then this packet would be retained in $Or_\rho$. Accordingly, rescue $Or_\rho$ means to deliver a packet in $Or_\rho$ until its loss-state didn't belong to $\Phi^i$. In this way, to recover $P_{\Phi^i}$ is equivalent to rescue $Or_{\Phi^i}$. Now, let us proof that the expect number of retransmissions required to rescue $Or_\rho$ ($\rho\in\Phi^i$) under the above definition is

$$\Omega'_{Or_\rho}=\frac{K\Pr\{\rho\}}{1-\prod_{j=i}^{N}\omega_j} \tag{24}$$

First, let $\rho_1$ denote the pattern that $\rho_1\in\Phi^i$ and $W(\rho_1)=i-1$, then Eq. (24) is obviously true for $\rho_{i-1}$. Then Let $\rho_2(t)$ denote the vector the $t^{th}$ entry is zero and $W(\rho_2(t))=i-2$. Based on Theorem 1, to rescue loss packets in $Or_{\rho_2(t)}$ relating to $\rho_2(t)$ (for simply, we also say that to rescue $\rho_2(t)$ in $Or_{\rho_2(t)}$), the expect number is

$$\Omega'_{\rho_2(t)}=\frac{K\Pr\{\rho_2(t)\}}{1-\omega_t\prod_{k=i}^{N}\omega_k}$$

and to rescue $\rho_1$ in $Or_{\rho_2(t)}$, the number is

$$\Omega'_{\rho_2(t)\to\rho_1}=\frac{\Omega'_{\rho_2(t)}(1-\omega_t)\prod_{k=i}^{N}\omega_k}{1-\prod_{k=i}^{N}\omega_k}$$

Since $\rho_2(t)$ can only be convert to $\rho_1$, the number required to rescue $Or_{\rho_2(t)}$ is

$$\begin{aligned}\Omega'_{Or_{\rho_2(t)}}&=\Omega'_{\rho_2(t)}+\Omega'_{\rho_2(t)\to\rho_1}\\&=\frac{K\Pr\{\rho_2(t)\}}{1-\prod_{k=i}^{N}\omega_k}\end{aligned}$$

Now, suppose, this thesis is true for any $\rho_{n-1}\in\Phi^i$, $W(\rho_{n-1})=i-n+1>0$. Then let $\rho_n(T)\in\Phi^i$, $T=\{t_1\dots,t_n\}$, denote the pattern the $T^{th}$ entries are zero and $W(\rho_n)=i-n$. Using Theorem 1, to rescue $\rho_n(T)$ in $Or_{\rho_n(T)}$, the number is

$$\Omega'_{\rho_n(T)}=\frac{K\Pr\{\rho_n(T)\}}{1-\prod_{j=1}^{n}\omega_{t_j}\prod_{k=i}^{N}\omega_k}$$

and

$$\Omega_{\rho_n(T)\to\rho_x(T')}=\Omega'_{\rho_n(T)}\Pr\{\rho_x(T')|\rho_n(T)\}$$

packets transformed to $\rho_x(T')\in\Phi^i$, $1\leqslant x<n$ $T'=\{t'_1,\dots,t'_x\}$. If we operate these packets as well as $Or_{\rho_x(T')}$, then in the light of assumption, the required number to rescue them is

$$\Omega'_{Or'_{\rho_x(T')}}=\frac{\Omega_{\rho_n(T)\to\rho_x(T')}}{1-\prod_{k=i}^{N}\omega_k}$$

Therefore, we have

$$\begin{aligned}\Omega'_{Or_{\rho_n(T)}}&=\sum_{\rho_x(T')\in\Phi^i}\Omega'_{Or'_{\rho_x(T')}}+\Omega'_{\rho_n(T)}\\&=\frac{\Omega'_{\rho_n(T)}}{1-\prod_{k=i}^{N}\omega_k}\cdot\sum_{\rho_x(T')\in\Phi^i}\Pr\{\rho_x|\rho_n\}+\Omega'_{\rho_n(T)}\\&=\frac{\Omega'_{\rho_n(T)}(1-\prod_{j=1}^{n}\omega_{t_j})\prod_{j=i}^{N}\omega_j}{1-\prod_{k=i}^{N}\omega_k}+\Omega'_{\rho_n(T)}\\&=\frac{K\Pr\{\rho_n(T)\}}{1-\prod_{k=i}^{N}\omega_k}\end{aligned}$$

By induction the proposition is proofed. Moreover,

$$\Omega_{\Phi^i}=\Omega'_{Or_{\Phi^i}}=\frac{K\sum_{\rho\in\Phi^i}\Pr\{\rho\}}{1-\prod_{k=i}^{N}\omega_k}=\frac{K\prod_{k=i}^{N}\omega_k}{1-\prod_{k=i}^{N}\omega_k}$$

then we figure out the expected number of retransmissions required to recover the entire loss packets,

$$\Omega_{all}=\sum_{i=1}^{N}\Omega_{U^i}=\sum_{i=1}^{N}\Omega_{\Phi^i}=\sum_{i=1}^{N}\frac{K\prod_{k=i}^{N}\omega_k}{1-\prod_{j=i}^{N}\omega_k} \tag{25}$$

Dividing $KN$, then gives the theorem. ■